\documentclass[a4paper]{elsart}
\usepackage{natbib}
\usepackage{amssymb}

\newcommand{\z}[0]{{\mathbf z}}
\newcommand{\ie}[0]{{i.e.}, }
\newcommand{\eg}[0]{{e.g.}, }
\usepackage{graphics}

\newcommand{\image}[4]
{
\begin{figure}[htb]
\centering
\makebox{
  \resizebox
      {#2cm}{!}
      {\includegraphics{#1}}
}
\caption{#4}
\label{#3}
\end{figure}
}
\def\astrobj#1{#1}

\begin{document}
\begin{frontmatter}

\title{A Kinematic Model for Gamma Ray Bursts and Symmetric Jets}

\author{M. Thulasidas\corauthref{cor}}
\corauth[cor]{Corresponding author. Phone: +65 6874 3123, FAX: +65 6774 4990}
\address{New Initiatives, Institute for Infocomm Research,
21 Heng Mui Keng Terrace, Singapore 119613.}
\ead{manoj@i2r.a-star.edu.sg}

\begin{abstract}
  Gamma ray bursts (GRB) occur at random points in the sky at
  cosmological distances.  They emit intense $\gamma$ rays for a brief
  period.  The spectrum then evolves through X--ray, optical region to
  possibly radio frequency.  Though there are some models, the origin
  and time evolution of GRB are not well understood.  Extragalactic
  radio sources also exhibit a baffling array of features that are
  poorly understood -- the core emission in ultraviolet region, lobes
  in RF range, transient $\gamma$ and X--ray emissions etc.  These two
  phenomena appear to be very different, but the time evolution of the
  core emission of radio sources is essentially the same as GRBs,
  though with different time constants.  Here, we present a model
  unifying GRB and roughly symmetric radio sources based on light
  travel time effect and superluminality.  The light travel time
  effect influences the way we perceive superluminal motion.  An
  object, moving across our field of vision at superluminal speeds,
  will appear to us as two objects receding from a single point.  The
  time evolution of the Doppler shifted radiation of such a
  superluminal object bears remarkable similarity to that of GRB and
  radio sources.  Based on these observations, we derive a kinematic
  model for radio sources and GRBs and explain the puzzling features
  listed above.  Furthermore, we predict the angular motion of the
  hotspots or knots in the jets and compare it to the proper motions
  reported in the literature \citep{M87}. We also compare the proper
  motion of a microquasar \citep{superluminal3} with our model and
  show excellent agreement.  Our model also explains the observed
  blue/UV spectrum (and its time evolution) of the core region and the
  RF spectrum of the lobes, and why the radio sources appear to be
  associated with galactic nuclei. We make other quantitative
  predictions, which can be verified.

\end{abstract}


\end{frontmatter}

\section{Introduction}

Gamma Ray Bursts (GRBs) are short and intense flashes of $\gamma$ rays
in the sky, lasting from a few milliseconds to several minutes
\citep{GRB1}.  They are characterized by a prompt emission and an
after glow at progressively softer energies. Thus, the initial
$\gamma$ rays are promptly replaced by X--rays, light and even radio
frequency waves. This softening of the spectrum has been known for
quite some time \citep{GRB5}.  More recently, an inverse decay of the
peak energy with varying time constant has been used to empirically
fit the observed time evolution of the peak energy \citep{GRB3,GRB4}.
According to the fireball model, GRBs are produced when the energy of
highly relativistic flows in stellar collapses are dissipated.  There
is an overall agreement with the observed phenomena, though the exact
nature of the structure of the radiation jets from the collapsing
stars is still an open question.  In this article, we present a
radically different model explaining the origin of GRBs, their
occurrence in seemingly random points in the sky and the time
evolution of their spectra.

Symmetric radio sources (galactic or extragalactic) may appear to be a
completely distinct phenomenon.  However, their cores show a similar
time evolution in the peak energy, but with a much larger time
constant.  Other similarities have begun to attract attention in the
recent years \citep{GRB7}.  Due to observational constraints, the time
evolution of the spectra of radio sources has not been studied in
great detail, but one available example is the jet in \astrobj{3C 273}
\citep{RF2UV}, which shows a softening of the spectrum remarkably
similar to GRBs.  Our model unifies these two phenomena and make
detailed predictions on the kinematics of such radio sources.  We will
first look at the apparent superluminal motion observed in some jets
and its traditional explanation.  We will then proceed to derive the
model for explaining the kinematics such symmetric radio sources and
apply it to GRBs.

Some of the radio sources show superluminal motion in the transverse
direction.  We can measure the transverse velocity of a celestial
object almost directly using angular measurements, which are
translated to a speed using its known (or estimated) distance from us.
In the past few decades, scientists have observed
\citep{M87,superluminal2} objects moving at apparent transverse
velocities significantly higher than the speed of light.  Some such
superluminal objects were detected within our own galaxy
\citep{superluminal1,superluminal3,superluminal4,GRS1915}.  The
special theory of relativity (SR) states that nothing can accelerate
past the speed of light \citep{einstein}.  Thus a direct measurement
of superluminal objects emanating from a single point (or a small
region) would be a violation of SR at a fundamental level.  However,
before proclaiming a contradiction based on an apparent superluminal
motion, one has to establish that it is not an artifact of the way one
perceives transverse velocities.  \citet{rees} offered an explanation
why such apparent superluminal motion is not in disagreement with SR,
even before the phenomenon was discovered.  When an object travels at
a high speed towards an observer, at a shallow angle with respect to
his line of sight, it can appear to possess superluminal speeds.
Thus, a measurement of superluminal transverse velocity by itself is
not an evidence against SR.  In this article, we re-examine this
explanation (also known also as the light travel time effect.)  We
will show that the apparent symmetry of the extragalactic radio
sources is inconsistent with this explanation.

\image{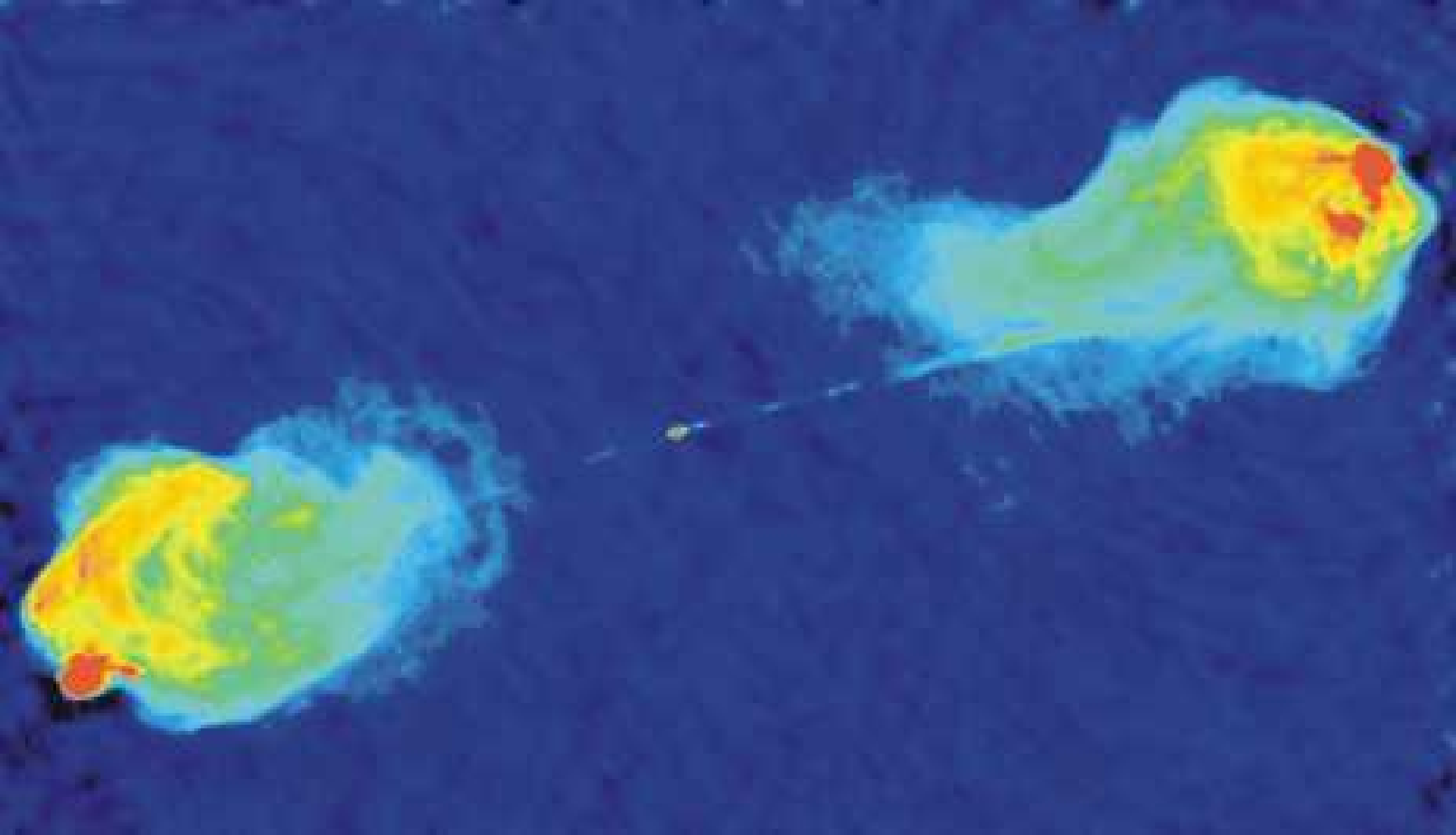}{8}{cyga} {False color image of the radio jet and lobes
  in the hyperluminous radio galaxy \astrobj{Cygnus A}. Red shows
  regions with the brightest radio emission, while blue shows regions
  of fainter emission.  Image courtesy of NRAO/AUI.}

Transverse superluminal motions are usually observed in quasars and
microquasars. Different classes of such objects associated with Active
Galactic Nuclei (AGN) were found in the last fifty years.
Fig.~\ref{cyga} shows the radio galaxy \astrobj{Cygnus A} \citep{cyga}
-- one of the brightest radio objects.  Many of its features are
common to most extragalactic radio sources -- the symmetric double
lobes, an indication of a core, an appearance of jets feeding the
lobes and the hotspots.  \citet{hotspot1} and \citet{hotspot2} have
reported more detailed kinematic features, such as the proper motion
of the hotspots in the lobes.  We will show that our perception of an
object crossing our field of vision at a constant superluminal speed
is remarkably similar to a pair of symmetric hotspots departing from a
fixed point with a decelerating rate of angular separation.  We will
make other quantitative predictions that can be verified, either from
the current data or with dedicated experiments.

\section{Traditional Explanation}

\image{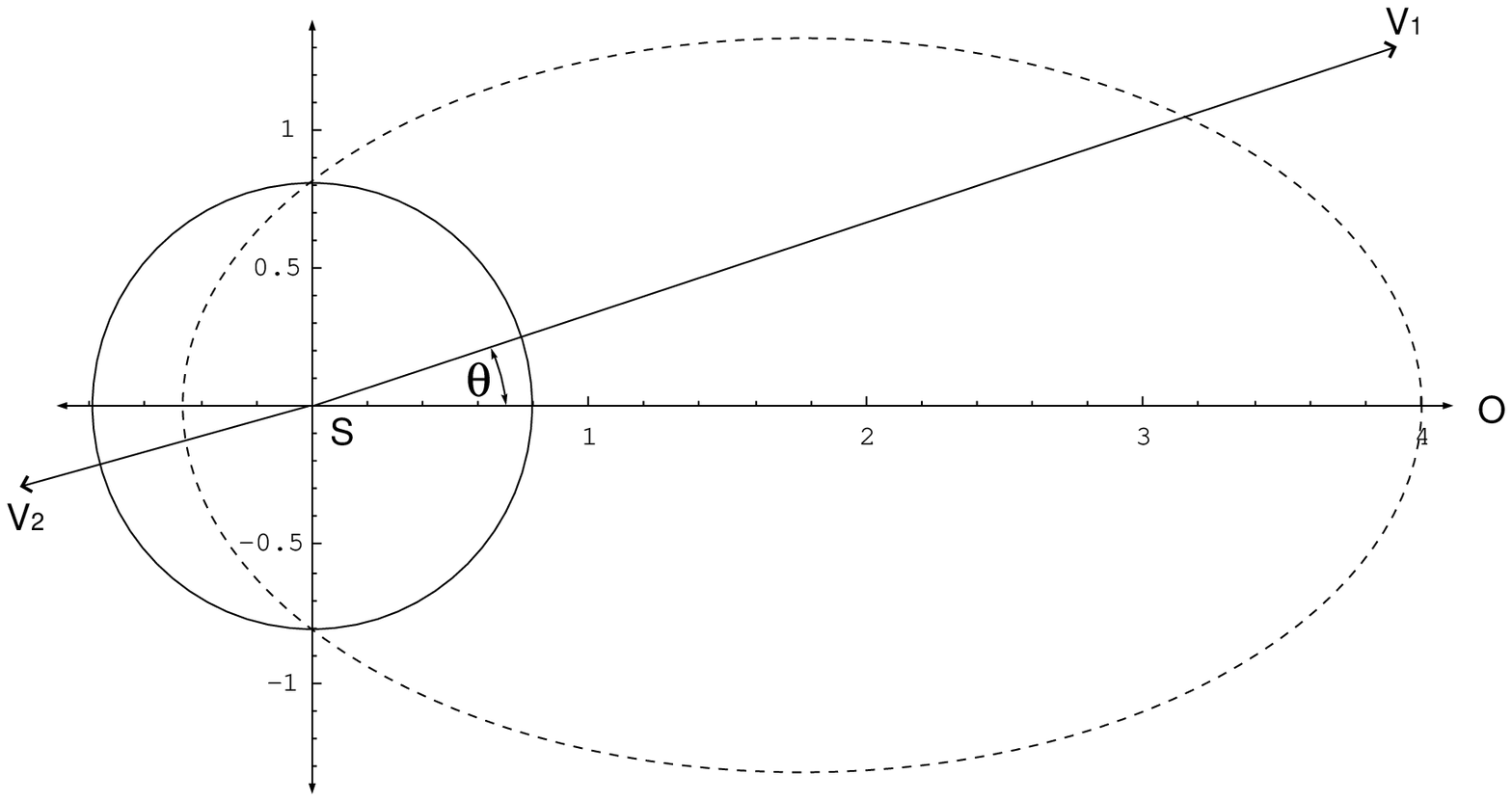}{8}{rees1} {Illustration of the traditional
  explanation for the apparent superluminal motion.  An object
  expanding at a speed $\beta = 0.8$, starting from a single point S.
  The solid circle represents the boundary one second later.  The
  observer is far away on the right hand side, O ($x\to\infty$).  The
  dashed ellipse is the apparent boundary of the object, as seen by
  the observer.}

First, we look at the traditional explanation of the apparent
superluminal motion by the light travel time effect.  Fig.~\ref{rees1}
illustrates the explanation of apparent superluminal motion as
described in the seminal paper by \citet{rees}.  In this figure, the
object at S is expanding radially at a constant speed of $0.8c$, a
highly relativistic speed.  The part of the object expanding along the
direction $V_1$, close to the line of sight of the observer, will
appear to be traveling much faster.  This will result in an apparent
transverse velocity that can be superluminal.

The apparent speed $\beta'$ of the object depends on the real speed
$\beta$ and the angle between its direction of motion and the
observer's line of sight, $\theta$.  As shown in
Appendix~\ref{sec.bp},
\begin{equation}
\beta' \quad=\quad \frac{\beta}{1\,-\,\beta \cos\theta} \label{eqn.1}
\end{equation}
Fig.~\ref{rees1} is a representation of equation~(\ref{eqn.1}) as
$\cos\theta$ is varied over its range.  It is the locus of $\beta'$
for a constant $\beta = 0.8$, plotted against the angle $\,\theta$.
The apparent speed is in complete agreement with what was predicted in
1966 (Fig.~1 in that article \citep{rees}).

For a narrow range of $\,\theta$, the transverse component of the
apparent velocity ($\,\beta'\sin\theta\,$) can appear superluminal.
From equation~(\ref{eqn.1}), it is easy to find this range:
\begin{equation}
\frac{1 - \sqrt{2 \beta^2-1}}{2\beta} < \cos\theta < \frac{1 +
  \sqrt{2 \beta^2-1}}{2\beta} \label{eqn.2}
\end{equation}
Thus, for appropriate values of $\,\beta (>\frac{1}{\sqrt{2}})\,$ and
$\,\theta\,$ (as given in equation~(\ref{eqn.2})), the transverse
velocity of an object can seem superluminal, even though the real
speed is in conformity with the special theory of relativity.

While equations~(\ref{eqn.1}) and (\ref{eqn.2}) explain the apparent
transverse superluminal motion the difficulty arises in the
recessional side.  Along directions such as $V_2$ in Fig.~\ref{rees1},
the apparent velocity is always smaller than the real velocity.  The
jets are believed to be emanating from the same AGN in opposite
directions.  Thus, if one jet is in the $\cos\theta$ range required
for the apparent superluminal motion (similar to $V_1$), then the
other jet has to be in a direction similar to $V_2$.  Along this
direction, the apparent speed is necessarily smaller than the real
speed, due to the same light travel time effect that explains the
apparent superluminal motion along $V_1$. Thus, the observed symmetry
of these objects is inconsistent with the explanation based on the
light travel time effect.  Specifically, superluminality can never be
observed in both the jets (which, indeed, has not been reported so
far).  However, there is significant evidence of near symmetric
outflows \citep{asymmetry} from a large number of objects similar to
the radio source in Fig.~\ref{cyga}.

One way out of this difficulty is to consider hypothetical
superluminal speeds for the objects making up the apparent jets.  Note
that allowing superluminal speeds is not in direct contradiction with
the special theory of relativity, which does not treat superluminality
at all.  The original derivation \citep{einstein} of the theory of
coordinate transformation is based on the definition of simultaneity
enforcing the constancy of the speed of light.  The synchronization of
clocks using light rays clearly cannot be done if the two frames are
moving with respect to each other at superluminal speeds.  All the
ensuing equations apply only to subluminal speeds.  It does not
necessarily preclude the possibility of superluminal motion. However,
an object starting from a fixed point and accelerating past the speed
of light is clear violation of SR.

Another consequence of the traditional explanation of the apparent
superluminal speed is that it is invariably associated with a blue
shift. As given in equation~(\ref{eqn.2}), the apparent transverse
superluminal speed can occur only in a narrow region of $\cos\theta$.
In this region, the longitudinal component of the velocity is always
towards the observer, leading to a blue shift.  The existence of blue
shift associated with all superluminal jets has not been confirmed
experimentally.  Quasars with redshifts have been observed with
associated superluminal jets.  Two examples are: quasars \astrobj{3C
  279} \citep{q3c279} with a redshift $\z = 0.536$ and \astrobj{3C
  216} with $\z = 0.67$ \citep{q3c216}.  However, the Doppler shift of
spectral lines applies only to normal matter, not if the jets are made
up of plasma, as currently believed.  Thus, the current model of jets,
made up of plasma collimated by a magnetic field originating from an
accretion disc, can accommodate the lack of blue shift.

\section{A Model for Double--lobed Radio Sources GRBs}

Accepting hypothetical superluminal speeds, we can clearly tackle the
second consequence of the traditional explanation (namely, the
necessity of blue shift along with apparent superluminal motion.)
However, it is not clear how we perceive superluminal motion, because
the light travel time effect will influence the way we perceive the
kinematics.  In this section, we will show that a single object moving
superluminally, in a transverse direction across our field of vision, will
look like two objects departing from a single point in a roughly
symmetric fashion.

\subsection{Symmetric Jets}

\image{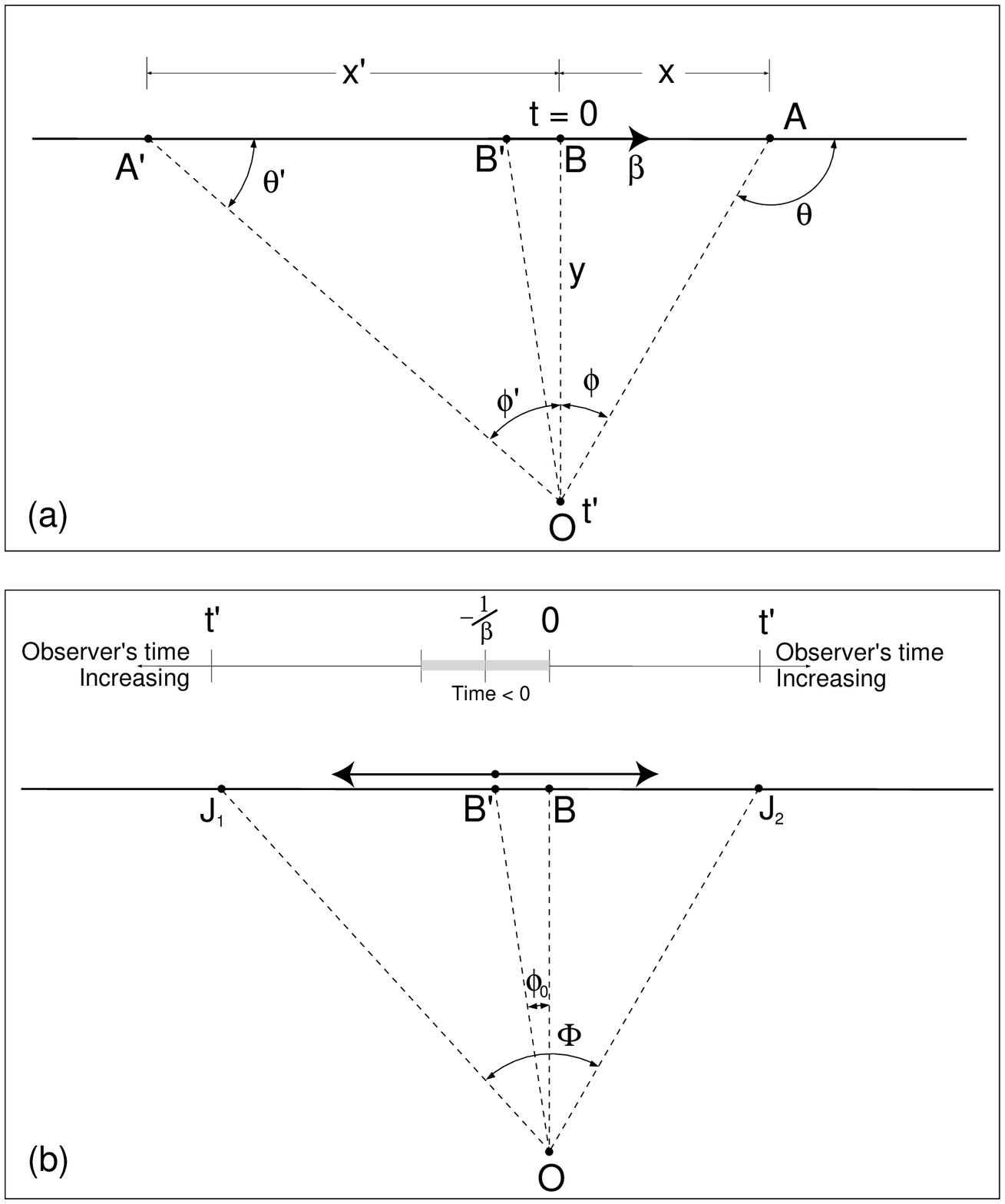}{8}{d} {The top panel (a) shows an object flying along
  $A'BA$ at a constant superluminal speed.  The observer is at $O$.
  The object crosses $B$ (the point of closest approach to $O$) at
  time $t=0$.  The bottom panel (b) shows how the object is perceived
  by the observer at $O$.  It first appears at $B'$, then splits into
  two.  The two apparent objects seem to go away from each other
  (along $J_1$ and $J_2$) as shown.}

Consider an object moving at a superluminal speed as shown in
Fig.~\ref{d}(a).  The point of closest approach is $B$.  At that
point, the object is at a distance of $y$ from the observer at $O$.
Since the speed is superluminal, the light emitted by the object at
some point $B'$ (before the point of closest approach $B$) reaches the
observer {\em before\/} the light emitted at $A'$.  This gives an
illusion of the object moving in the direction from $B'$ to $A'$,
while in reality it is moving in the opposite direction.

We use the variable $t'$ to denote the the observer's time.  Note
that, by definition, the origin in the observer's time axis is set
when the object appears at $B$.  $\phi$ is the observed angle with
respect to the point of closest approach $B$. $\phi$ is defined as
$\theta - \pi/2$ where $\theta$ is the angle between the object's
velocity and the observer's line of sight.  $\phi$ is negative for
negative time $t$.

It is easy to derive the relation between $t'$ and $\phi$.  (See
Appendix~\ref{sec.super} for the mathematical details.)
\begin{equation}
 t' = y\left( \frac{\tan\phi}\beta + \frac{1}{\cos\phi} - 1\right)
 \label{eqn.3}
\end{equation}
Here, we have chosen units such that $c = 1$, so that $y$ is also the
time light takes to traverse $BO$.  The observer's time is measured
with respect to $y$.  \ie $t' = 0$ when the light from the point of
closest approach $B$ reaches the observer.

The actual plot of $\phi$ as a function of the observer's time is
given in Fig.~\ref{PhiVsTp0}.  Note that for for subluminal speeds,
there is only one angular position for any given $t'$.  The time axis
scales with $y$.  For subluminal objects, the observed angular
position changes almost linearly with the observed time, while for
superluminal objects, the change is parabolic.

Equation~(\ref{eqn.3}) can be approximated using a Taylor series
expansion as:
\begin{equation}
t' \approx y\left(\frac\phi\beta +
  \frac{\phi^2}{2}\right)\label{eqn.4}
\end{equation}
From the quadratic equation~(\ref{eqn.4}), one can easily see that the
minimum value of $t'$ is $t'_{\rm min} = -y/2\beta^2$ and it occurs at
$\phi_{0}=-1/\beta$.  Thus, to the observer, the object first appears
(as though out of nowhere) at the position $\phi_0$ at time $t'_{\rm
  min}$.  Then it appears to stretch and split, rapidly at first, and
slowing down later.  This apparent time evolution of the object is
shown in Fig.~\ref{grs}, where it is compared to the microquasar
\astrobj{GRS 1915+105} \citep{superluminal1,GRS1915}.

The angular separation between the objects flying away from each other
is:
\begin{equation}
 \Phi = \frac{2}{\beta}\sqrt{1+\frac{2\beta^2}{y}t'} =
 \frac{2}{\beta}\left(1+\beta\phi\right)
\end{equation}
And the rate at which the separation occurs is:
\begin{equation}
 \frac{d\Phi}{dt'} = \sqrt{\frac{2}{y t_{\rm age}}} =
 \frac{2\beta}{y\left(1+\beta\phi\right)} \label{eqn.5}
\end{equation}
where $ t_{\rm age} = t' - t'_{\rm min}$, the apparent age of the
symmetric object.

\image{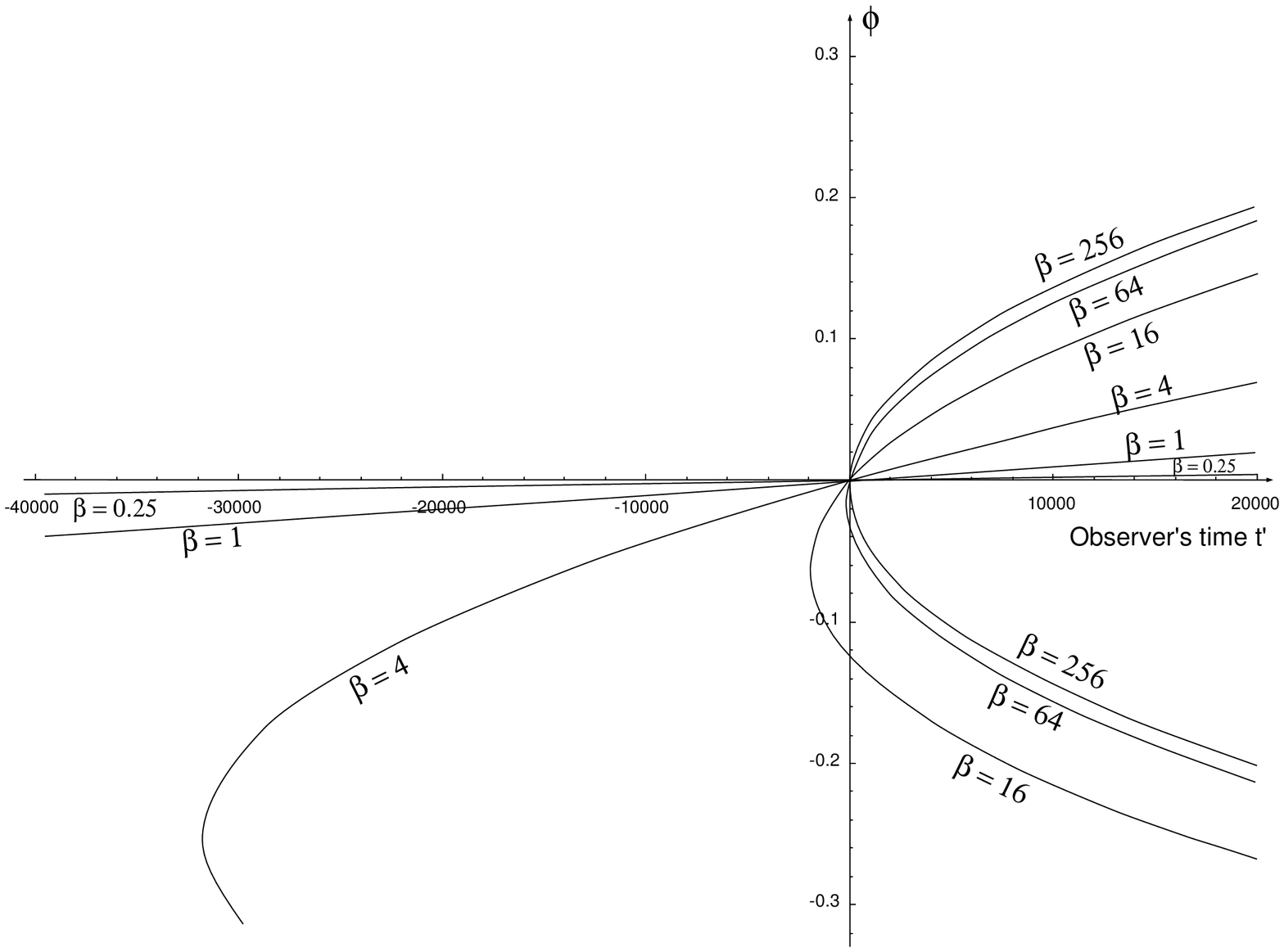}{8}{PhiVsTp0} {The apparent angular positions of an
  object traveling at different speeds at a distance $y$ of one
  million light years from us.  The angular positions ($\phi$ in
  radians) are plotted against the observer's time $t'$ in years.}

\image{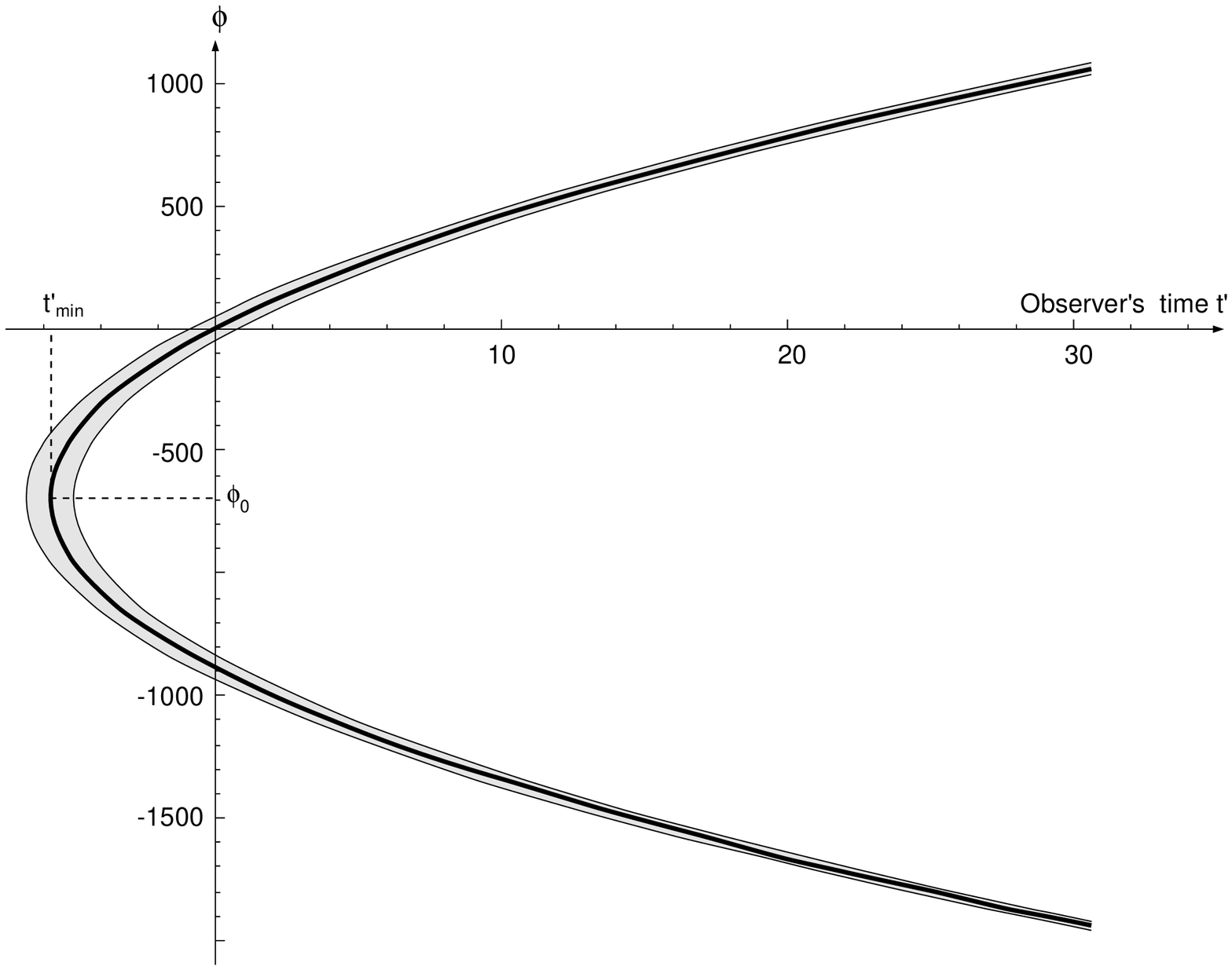}{8}{PhiVsTp} {The apparent angular positions and sizes
  of an object traveling at $\beta = 300$ at a distance $y$ of one
  million light years from us.  The angular positions ($\phi$ in arc
  seconds) are plotted against the observer's time $t'$ in years. The
  shaded region represents the apparent angular spread of the object,
  with an assumed diameter of 500 light years.}

This discussion shows that a single object moving across our field of
vision at superluminal speed creates an illusion of an object
appearing at a at a certain point in time, stretching and splitting
and then moving away from each other.  This time evolution is given in
equation~(\ref{eqn.3}), and illustrated in the bottom panel of
Fig.~\ref{d}(b).  Note that the apparent time $t'$ is reversed with
respect to the real time $t$ in the region $A'$ to $B'$.  An event
that happens near $B'$ appears to happen before an event near $A'$.
Thus, the observer may see an apparent violation of causality, but it
is just a part of the light travel time effect.

Fig.~\ref{PhiVsTp} shows the apparent width of a superluminal object
as it evolves.  The width decreases with time, along its direction of
motion.  (See Appendix~\ref{sec.dx} for the mathematical details.)
Thus, the appearance is that of two spherical objects appearing out of
nowhere, moving away from each other, and slowly getting compressed
into thinner and thinner ellipsoids and then almost disappearing.

If there are multiple objects, moving as a group, at roughly constant
superluminal speed along the same direction, their appearance would be
a series of objects appearing at the same angular position and moving
away from each other sequentially, one after another.  The apparent
knot in one of the jets always has a corresponding knot in the other
jet.

The calculation presented in this article is done in two dimensions.
If we generalize to three dimensions, we can explain the precession
observed in some systems.  Imagine a cluster of objects, roughly in a
planar configuration (like a spiral galaxy, for instance) moving
together at superluminal speeds with respect to us.  All these objects
will have the points of closest approach to us in small angular region
in our field of vision -- this region is around the point of minimum
distance between the plane and our position.  If the cluster is
rotating (at a slow rate compared to the superluminal linear motion),
then the appearance to us would be the apparent jets changing
directions as a function of time.  The exact nature of the apparent
precession depends on the spatial configuration of the cluster and its
angular speed.

If we can measure the angle $\phi_0$ between the apparent core and the
point of closest approach, we can directly estimate the real speed of
the object $\beta$.  We can clearly see the angular position of the
core.  However, the point of closest approach is not so obvious.  We
will show in the next section that the point of closest approach
corresponds to zero redshift.  (This is obvious intuitively, because
at the the point of closest approach, the longitudinal component of
the velocity is zero.)  If this point ($\phi_0$) can be estimated
accurately, then we can measure the speed directly, from the relation
$\phi_0 = -1/\beta$.

\subsection{Redshifts of the Hotspots}
In the previous section, we showed how a superluminal object can
appear as two objects receding from a core.  We can also work out the
time evolution of the redshift of the two apparent objects (or
hotspots).  However, as relativistic Doppler shift equation is not
defined for superluminal speeds, we need to work out the relationship
between the redshift ($\z$) and the speed ($\beta$) from first
principles.  This is easily done (see Appendix~\ref{sec.z} for the
mathematical details):
\begin{equation}
1 \,+\, \z \,=\, \left|1 \,-\, \beta\cos\theta\right| \,=\, \left|1
  \,+\, \beta\sin\phi\right|
\end{equation}

Since we allow superluminal speeds in our model of extragalactic radio
sources, we can explain the radio frequency spectra of the hotspots as
extremely red-shifted blackbody radiation.  The $\beta$s involved in
this explanation are typically very large, and we can approximate the
redshift as:
\begin{equation}
1 \,+\, \z \,\approx\,  \left|\beta\phi\right| \,\approx\,
\frac{\left|\beta\Phi\right|}{2}
\end{equation}
Assuming the object to be a black body similar to the sun, we can
predict the peak wavelength (defined as the wavelength at which the
luminosity is a maximum) of the hotspots as:
\begin{equation}
\lambda_{\rm max} \,\approx\, (1+\z) 480nm \,\approx\,
\frac{\left|\beta\Phi\right|}{2} 480nm \label{eqn.z1}
\end{equation}
where $\Phi$ is the angular separation between the two hotspots.

This shows that the peak RF wavelength increases linearly with the
angular separation.  If multiple hotspots can be located in a twin jet
system, their peak wavelengths will depend only on their angular
separation, in a linear fashion.  The variation of the emission
frequency from ultraviolet to RF as $\phi$ increases along the jet is
clearly seen in the photometry of the jet in \astrobj{3C 273}
\citep{RF2UV}.  Furthermore, if the measurement is done at a single
radio frequency, intensity variation can be expected as the hotspot
moves along the jet.  Note that the limiting value of $|1+\z|$ is
roughly $\beta$, which gives an indication of the hyperluminal speeds
required to push the black body radiation to RF spectra.

\subsection{Time Evolution of GRB spectra}

\image{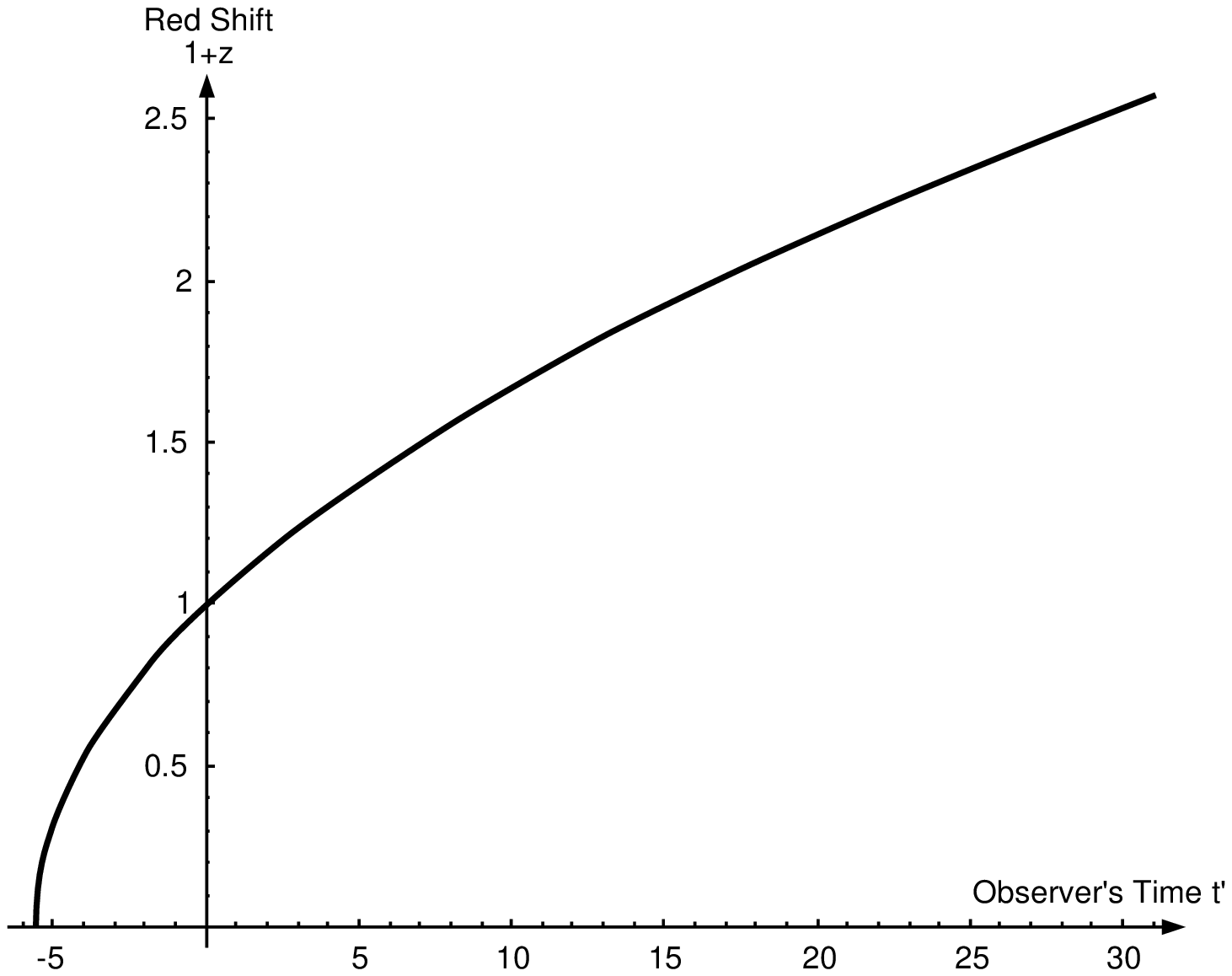}{8}{zVsTp}{Time evolution of the redshift from a
  superluminal object.  It shows the redshifts
  expected from an object moving at $\beta = 300$ at a distance of ten
  million light years from us.  The X axis is the observer's time in
  years.  (Since the X axis scales with time, it is also the redshift
  from an object at 116 light days -- ten million light seconds --
  with the X axis representing $t'$ in seconds.)}

The evolution of redshift of the thermal spectrum of a hyperluminal
object also holds the explanation for $\gamma$ ray bursts.  Since we
know the dependence of the observer's time $t'$ and the redshift
$1+\z$ on the real time $t$, we can plot one against the other
parametrically.  (See Appendix~\ref{sec.dz}.) Fig.~\ref{zVsTp} shows
the variation of redshift as a function of the observer's time
($t'$).  It shows that the observed spectra of a superluminal object
is expected to start at the observer's time $t'_{\rm min}$ with heavy
(infinite) blue shift.  The spectrum of the object rapidly softens and
soon evolves to zero redshift and on to higher values.  The rate of
softening depending on the its speed and distance from us, which is
the only difference between GRBs and symmetric jets (or radio sources)
in this model.  Note that to the observer, before $t'_{\rm min}$,
there is no object.  In other words, there is a definite point in the
observer's time when the GRB is ``born'', with no indication of its
impending birth before that time.  This birth does not correspond to
any cataclysmic event (as would be required in the collapsar/hypernova
or the ``fireball'' model) at the distant object.  It is just an
artifact of our perception.

There are two solutions for $\z$ for any given $t'$, corresponding to
the apparent objects at the two different positions.  However, as
shown in Appendix~\ref{sec.dz}, they are nearly identical (especially
for small $\phi$).  For $t' < 0$, there is a strong blue shift, which
explains the observed, transient hard X--ray spectra of some of the
symmetric jets.  \citet{xray1} have a recent survey of such data,
though the currently favored explanation for such transient emissions
is a stellar tidal disruption scenario.  The small difference between
the redshifts of the two apparent objects may explain the double peak
structure observed in the spectral data of some of the AGNs
\citep{doublepeak}.

Note that the X axis in Fig.~\ref{zVsTp} scales with time.  We have
plotted an object with $\beta = 300$ and $y =$ ten million light
years, with X axis is $t'$ in years.  It is also the variation of
$1+\z$ for an object at $y =$ ten million light seconds (or 116 light
days) with X axis in seconds.  The former corresponds to a symmetric
jets and the latter to a GRB.  Thus, for a GRB, the spectral evolution
takes place at a much faster pace.  Clearly, different combinations of
$\beta$ and $y$ can be fitted to describe different GRB spectral
evolutions.

We can also eliminate $t$ and study the time evolution of $1+\z$. In
order to make the algebra more manageable, we define $\tau = y/\beta$,
a characteristic time scale for the GRB (or the radio source).  This
is the time the object would take to reach us, if it were coming directly
toward us.  We also define the age of the GRB (or radio source) as $=
t_{\rm age} = t' - t'_{\rm min}$.  This is just the observer's time
($t'$) shifted by the time at which the object first appears to him
($t'_{\rm min}$).  With these notations (and for small values $t$), it
is possible to write the time dependence of $\z$ as:
\begin{equation}
  \label{eqn.z}
1 + \z = \left| {1 + \frac{{\beta ^2 \left( { - \tau  \pm \sqrt
            {2 \beta  t_{{\rm age}}} } \right)}}{{\beta t_{{\rm age}}
        + \tau /2 \mp \sqrt {2 \beta t_{{\rm age}}}  + \beta ^2 \tau
      }}} \right|
\end{equation}
for small values of $t \ll \tau$.  (The derivation of this equation
can be found in Appendix~\ref{sec.dz}.)

Since the peak energy of the spectrum is inversely proportional to the
redshift, it can be written as
\begin{equation}
  \label{eqn.Epk0}
  E_{\rm pk}(t_{{\rm age}}) = \frac{ E_{\rm pk}(t_{\rm min})}{1 + C_1
  \,\sqrt{\frac{t_{{\rm age}}}{\tau}} + C_2\,\frac{t_{{\rm age}}}{\tau}}
\end{equation}
where $C_1$ and $C_2$ are coefficients to be estimated by the Taylor
series expansion of equation~(\ref{eqn.z}) or by fitting.

The evolution of the peak energy ($E_{\rm pk}(t)$) has been empirically
modeled \citep{GRB6} as:
\begin{equation}
  \label{eqn.Epk}
  E_{\rm pk}(t) = \frac{ E_{\rm pk,0}}{(1+t/\tau)^\delta}
\end{equation}
where $t$ is the time elapsed after the onset ($= t_{\rm age}$ in our
notation), $\tau$ is a time constant and $\delta$ is the hardness
intensity correlation (HIC).  Out of the seven fitted values of
$\delta$ from \citet{GRB6}, one can calculate the average as $\delta =
1.038\pm0.014$, with the individual values ranging from $0.4$ to
$1.1$.  A fit to equation~(\ref{eqn.Epk0}) may give better results,
but within the statistics, it may not rule out or validate either
model.  Furthermore, it is not an easy fit, because there are too
many unknowns.  However, it is easy to see that the shape of
equations~(\ref{eqn.Epk0}) and (\ref{eqn.Epk}) are remarkably similar.

\subsection{Summary of Predictions}
Some of the different quantitative predictions of the model are
recapitulated here.  These are predictions that are relatively easy to
verify with existing data.
\begin{itemize}
\item The appearance of a single object moving across our field of
  vision at superluminal speed is that of an object appearing at a
  point, splitting and moving away in opposite directions.
\item The core will always have a fixed angular position.
\item The new superluminal knots appearing in the jets will always
  appear in pairs.
\item The two apparent objects will shrink monotonically. As the knots
  move away from the core, they become thinner and thinner ellipsoids,
  contracting along the direction of motion while the transverse size
  remains roughly constant.
\item The separation speed is very high in the beginning, but it slows
  down parabolically with time.
\item The hotspots have almost identical RF spectra (and redshifts).
\item The RF wavelength at which the luminosity of the hotspots is a
  maximum increases linearly with the angular separation between them.
\item Close to the core, the the spectrum is heavily blue shifted.
  Thus, the object can be a strong X--ray or even $\gamma$ ray source
  for a brief period of time.  After that, the spectrum moves through
  optical to RF region.
\item Since GRBs and symmetric jets are essentially the same
  cosmological phenomenon in this model, at least some of the GRBs
  will evolve to be symmetric jets with possible superluminal
  transverse speeds.
\end{itemize}

A clear indication of a movement in the core's angular position, or a
superluminal knot appearing without a counterpart in the opposite jet
will be strong evidence against our model based on superluminality.
On the other hand, a clear measurement of apparent superluminal motion
in both the jets (not reported so far) will provide a convincing
indication that the conventional explanation is inadequate.

\subsection{Comparison to Measurements}
\image{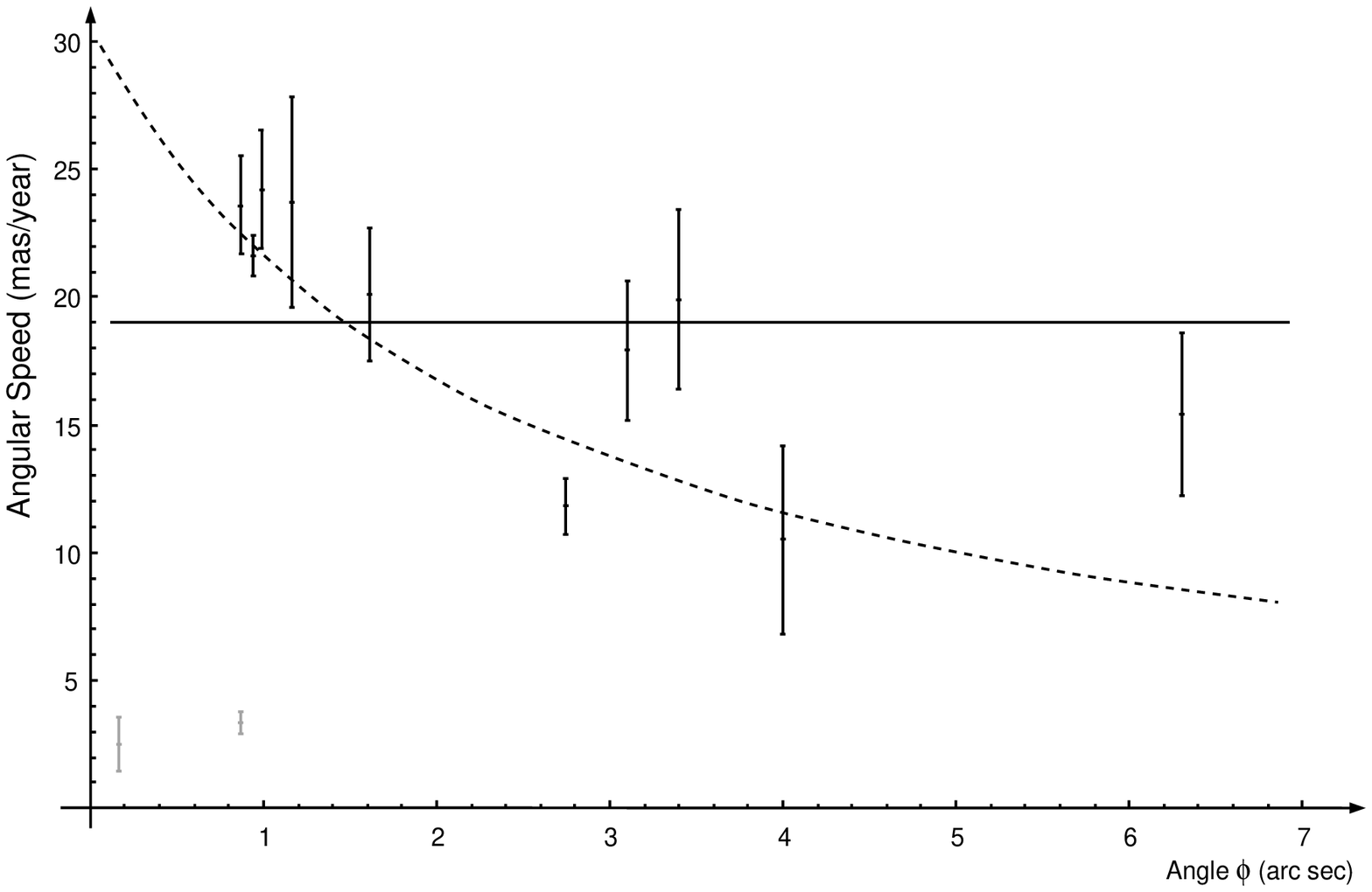}{8}{dphi} {The angular speed of M87 (as a function of
  the feature angle $\phi$) fitted against our model. The solid line
  constrains the distance of closest approach to the estimated
  distance of M87 -- 52 million light years, giving a $\beta = 4.8$.
  The dotted line is a free fit giving $\beta = 84\,000$.}

\citet{M87} have reported proper motion in one of the jets of
\astrobj{M87} as a function of the angle ($\phi$) between the apparent
core and the feature points.  \astrobj{M87} is estimated to be at
about 52 million light years away from us, which gives us the value of
$y$. In equation~(\ref{eqn.5}), we have the apparent angular speed
($d\Phi/dt'$) as a function of $\phi$.  Making the reasonable
approximation $\Phi \approx 2\phi$, we can fit these data to our
equation.  The result is shown in Fig.~\ref{dphi}.  The fit gives a
value $\beta = 4.8$.  A cluster of objects flying across our field of
vision, about five times faster than light and at a distance of 52
million light years, will look like two jets moving away from each
other at roughly 38 mas/year.  If one of the two jets is hidden for
some reason, the appearance will be a single jet of objects moving
away from a point with an angular speed of about 19 mas/year.  Note
that we exclude the first two points from the fit.  In this region
close to the core, the appearance of new objects makes it difficult to
track the features.

\image{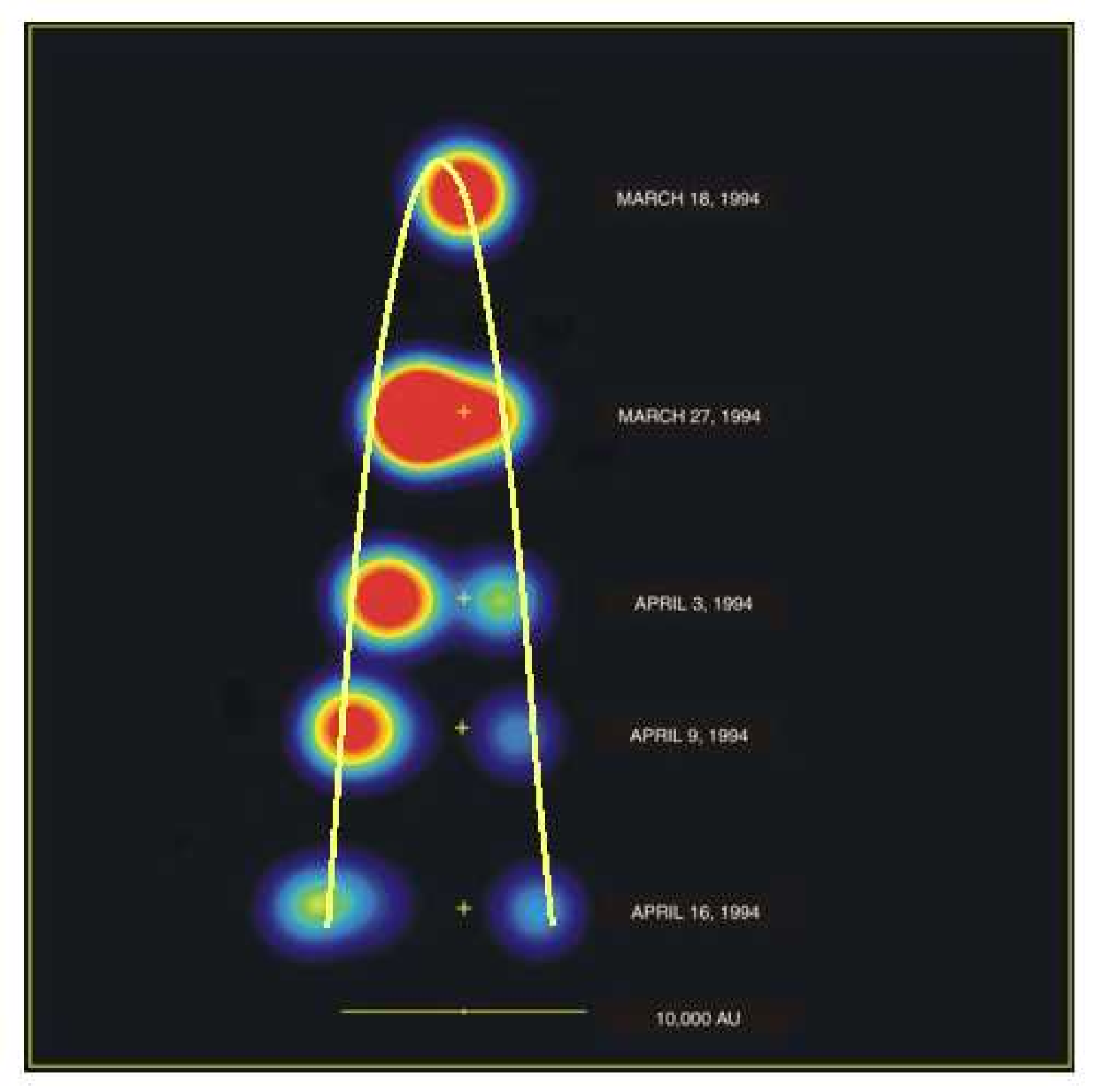}{8}{grs} {Fit of our model to the time evolution
  measurements of \astrobj{GRS1915+105}. The yellow curve overlaid
  corresponds to our perception of a single superluminal object,
  traveling at $\beta = 3\,000$ across our field of vision.}

A much better fit can be obtained if we were to let the distance $y$
also float.  The resulting $\beta$ of about 90\,000 may explain the
spectra of the hotspots. While the estimated $\beta$ may look
excessive, once superluminal motion is allowed, there is no a priori
reason why it should not take any value at all.  Fig.~\ref{grs} shows
another comparison of our model to the data available in the
literature.  Here, the time evolution of the microquasar \astrobj{GRS
  1915+105} \citep{superluminal1,GRS1915} is fitted to our model of a
single superluminal object. The deceleration of relativistic jets (one
of the predictions from our model) has been observed in the
Microquasar \astrobj{XTE J1550-564} \citep{sci_microquasar}, though it
is currently believed to be an effect similar to frictional drag.

AGNs are known to have intensely blue or ultraviolet core, not easily
explained by thermal models.  But, this is an expected feature in our
model. As seen in equation~(\ref{eqn.z1}), the core (where $\Phi\to0$)
must have a highly blue shifted spectrum. A clear evolution of
emission frequency from ultraviolet to RF is seen in the photometry of
the jet in \astrobj{3C 273} \citep{RF2UV}.  The spectrum shifts from
lower wavelength to higher as a function of the angular distance from
the core, strikingly consistent with our prediction.

This shifting of peak frequency can be seen at a much larger scale in
Fig.~\ref{cyga}. The size of the optical core is about a tenth of the
angular separation between the hotspots.  If we model \astrobj{Cygnus
  A} as a collection of objects moving together at superluminal
speeds, the core region would have emissions in the $\gamma$, X-ray,
UV or optical region.  As we move away towards the hotspots, the peak
frequency would continuously shift to RF.  This behavior is indeed
reported \citep{cyga1} recently.  This also partially explains why
extragalactic radio sources seem to be associated with galactic
nuclei, instead of appearing randomly in the sky.  A large collection
of objects moving together (a large spiral galaxy, viewed from the
side, for instance) superluminally gives the impression of a smaller
stationary object with optical emission.  The apparent object is
likely to appear elongated along the direction of motion (with the
major axis along the direction of the jets), with RF lobes appearing
symmetrically farther away from the core.  If the motion is not along
a linear trajectory, we may see curved jets.

\section{Conclusions}
In this article, we presented a unified model for Gamma Ray Bursts and
jet like radio sources based on bulk superluminal motion.  We showed
that a single superluminal object flying across our field of vision
would appear to us as the symmetric separation of two objects from a
fixed core.  Using this fact as the model for symmetric jets and GRBs,
we explain their kinematic features quantitatively.  In particular, we
showed that the angle of separation of the hotspots is parabolic in
time, and the redshifts of the two hotspots are almost identical to
each other.  Even the fact that the spectra of the hotspots are in the
radio frequency region is explained by assuming hyperluminal motion
and the consequent redshift of the blackbody radiation.  The time
evolution of the black body radiation of a superluminal object is
extremely consistent with the softening of the spectra observed in
GRBs and symmetric jets. In addition, our model explains why there is
significant blue shift at the core regions of radio sources, why radio
sources seem to associated with optical galaxies and why GRBs appear
at random points with no advance indication of their impending
appearance.

The currently favored models for these phenomena require either
cataclysmic events (explosions and shock waves in the fireball model
of GRBs) or space singularities (black hole accretion for jets).  GRBs
are observed at a rate of one per day, and are expected to be observed
much more frequently with the Swift mission.  The number of observed
jet like sources observed also is on the increase.  This makes the
explanations based on singularities and explosions less appealing.
Our model presents a more attractive option based on how we perceive
superluminal motion.  However, it does not address the energetics
issues -- the origin of superluminality.

We presented a set of predictions and compared them to existing data.
The features such as the blueness of the core, symmetry of the lobes,
the transient $\gamma$ and X-Ray bursts, the measured evolution of the
spectra along the jet all find natural and simple explanations in this
model.  Note that our model does not preclude plasma jets that may be
related to space-time singularities or other massive objects and the
associated accretion discs.  The conventional explanation of the
apparent superluminal motion in asymmetric jets (\eg quasar
\astrobj{3C 279} \citet{q3c279}) also stands.  In fact, our model is
just a generalization of the conventional explanation.

We also explored the full implications of the traditional explanation
for the apparent superluminal motion observed in certain quasars and
microquasars.  The equation that explains the apparent superluminal
speeds predicts that objects receding from us should appear to be
moving slower.  Thus, in a symmetric radio sources where it is
observed, the superluminal motion can appear only in one of the jets.
The observed symmetry of these extragalactic radio sources (even
subluminal ones) is incompatible with the explanation.  Another
consequence is that an apparent superluminal motion (if the moving
objects are composed of normal matter rather than plasma) must always
show a blue shift, a redshifted object can never be superluminal.  In
our model, the requirement that an apparent superluminal motion be
associated with a blue shift does not apply any more.  Furthermore,
the jets are expected to be fairly symmetric.

We argued that superluminal motion is not inconsistent with the
special theory of relativity, which just does not deal with it.
Acceptance of superluminality has far-reaching consequences in other
long established notions of our universe. (See
Appendix~\ref{sec.universe} for an incomplete list.)  The description
of extragalactic (or galactic) radio sources in terms of superluminal
motion has a direct impact on our understanding of black holes.

\appendix
\newpage
\centerline{\large\bf Appendix}
\section{Mathematical Details}
\image{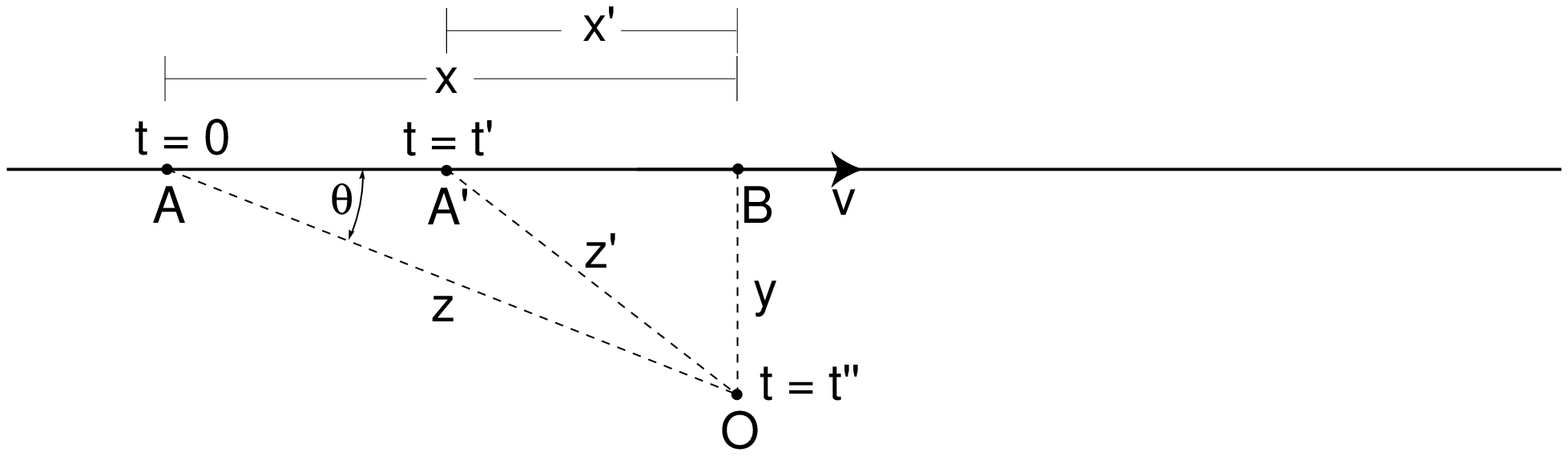}{8}{c} {The object is flying along $AB$, the observer
  is at $O$.  The object crosses $A$ at time $t=0$.  It reaches $A'$
  at time $t = t'$.  A photon emitted at $A$ reaches $O$ at time $t =
  t_0$, and a photon emitted at $A'$ reaches $O$ at time $t = t''$.}

\subsection{Velocity Profile of an Expanding Object}
\label{sec.bp}
In this section, we derive the ellipse in Fig.~\ref{rees1} from first
principles.  In Fig.~\ref{c}, there is an observer at $O$. An object
is flying by at a high speed $v = \beta c$ along the horizontal line
$AB$.  With no loss of generality, we can assume that $t = 0$ when the
object is at $A$. It passes $A'$ at time $t'$. The photon emitted at
time $t = 0$ reaches the observer at time $t = t_0$, and the photon
emitted at $A'$ (at time $t = t'$) reaches him at time $t = t''$.  The
angle between the object's velocity at $A$ and the observer's line of
sight is $\theta$.  We have the Pythagoras equations:
\begin{eqnarray}
z^2 \,&=&\, x^2 \,+\, y^2 \\
z'^2 \,&=&\, x'^2 \,+\, y^2\\
\Rightarrow\qquad \frac{x+x'}{z+z'} &\,=\,& \frac{z-z'}{x-x'} \label{eqn.6}
\end{eqnarray}
If we assume that $x$ and $z$ (distances at time $t_0$) are not very
different from $x'$ and $z'$ respectively (distances at time $t'$), we
can write,
\begin{equation}
\cos\theta = \frac{x}{z} \approx \frac{x+x'}{z+z'} =
\frac{z-z'}{x-x'}
\end{equation}
We define the real speed of the object as:
\begin{equation}
v \,=\, \beta\,c \,=\, \frac{x\,-\,x'}{t'}
\end{equation}
But the speed it {\em appears\/} to have will depend on when the
observer senses the object at $A$ and $A'$.  The apparent speed of the
object is:
\begin{equation}
v' \,=\, \beta'\,c \,=\, \frac{x \,-\, x'}{t'' \,-\,
t_0}
\end{equation}
Thus,
\begin{eqnarray}
\frac{\beta}{\beta'} \,&=&\, \frac{t''-t_0}{t'} \\
\,&=&\, 1 + \frac{z'-z}{ct'} \\
\,&=&\, 1 - \frac{x-x'}{ct'}\cos\theta \\
\,&=&\, 1 - \beta\cos\theta
\end{eqnarray}
which gives,
\begin{equation}
\beta' \quad=\quad
\frac{\beta}{1\,-\,\beta \cos\theta}\label{eqn.7}
\end{equation}

Fig.~\ref{rees1} is the locus of $\beta'$ for a constant $\beta =
0.8$, plotted against the angle $\,\theta$.

\subsection{Superluminal Redshift}
\label{sec.z}
Redshift ($\z$) defined as:
\begin{equation}
1 \,+\, \z \,=\, \frac{\lambda'}{\lambda}
\end{equation}
where $\lambda'$ is the measured wavelength and $\lambda$ is the known
wavelength.  In Fig.~\ref{c}, the number of wave cycles created in
time $t'$ between $A$ and $A'$ is the same as the number of wave
cycles sensed at $O$ between $t_0$ and $t''$.  Substituting the
values, we get:
\begin{equation} \frac{t'\, c}{\lambda} \,=\, {\frac{(t''\,-\,t_0)\,c}
{\lambda'}}
\end{equation}
Using the definitions of the real and apparent speeds, it is easy to get
\begin{equation}
\frac{\lambda'}{\lambda} \,=\, \frac{\beta}{\beta'}
\end{equation}
Using the relationship between the real speed $\beta$ and the
apparent speed $\beta'$ (equation~(\ref{eqn.7})),
we get
\begin{equation}
1 \,+\, \z \,=\, \frac{1}{1 \,+\, \beta'\cos\theta} \,=\, 1 \,-\,
\beta\cos\theta
\end{equation}
As expected, $\z$ depends on the longitudinal component of the
velocity of the object.  Since we allow superluminal speeds in this
calculation, we need to generalize this equation for $\z$ noting that
the ratio of wavelengths is positive.  Taking this into account, we
get:
\begin{equation}1 \,+\, \z \,=\, \left|\frac{1}{1 \,+\,
      \beta'\cos\theta}\right|
\,=\, \left|1 \,-\, \beta\cos\theta\right| \label{eqn.8}
\end{equation}

\subsection{Kinematics of Superluminal Objects}
\label{sec.super}
\image{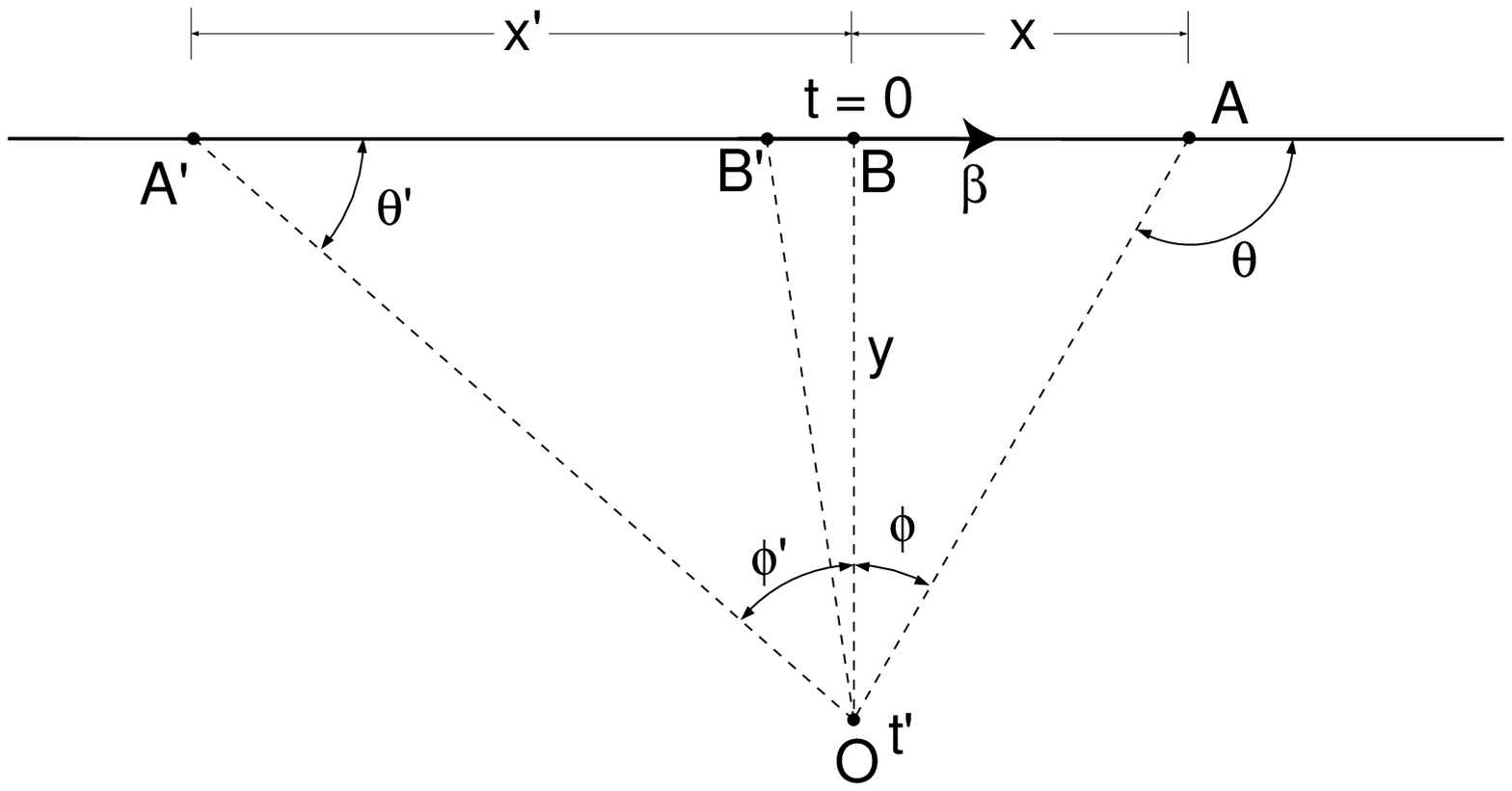}{8}{dd} {An object flying along $A'BA$ at a constant
  superluminal speed.  The observer is at $O$.  The object crosses $B$
  (the point of closest approach to $O$) at time $t=0$.}

The derivation of the kinematics is based on Fig.~\ref{dd}.  Here, an
object is moving at a superluminal speed along $A'BA$. At the point of
closest approach, $B$, the object is a distance of $y$ from the
observer at $O$. Since the speed is superluminal, the light emitted by
the object at some point $B'$ (before the point of closest approach
$B$) reaches the observer {\em before\/} the light emitted at $A'$.
This gives an illusion of the object moving in the direction from $B'$
to $A'$, while in reality it is moving from $A'$ to $B'$.  $\phi$ is
the observed angle with respect to the point of closest approach $B$.
$\phi$ is defined as $\theta - \pi/2$ where $\theta$ is the angle
between the object's velocity and the observer's line of sight.
$\phi$ is negative for negative time $t$.

We choose units such that $c = 1$, in order to make algebra simpler.
$t'$ denotes the the observer's time.  Note that, by definition, the
origin in the observer's time, $t'$ is set when the object appears at
$B$.

The real position of the object at any time $t$ is:
\begin{equation}
 x = y\tan\phi = \beta t
\end{equation}
A photon emitted at $t$ will reach $O$ after traversing the hypotenuse.
A photon emitted at $B$ will reach the observer at $t = y$, since we
have chosen $c = 1$.  If we define the observer's time $t'$ such that
the time of arrival is $t = t' + y$, then we have:
\begin{equation}
 t' = t + \frac{y}{\cos\phi} - y
\end{equation}
which gives the relation between $t'$ and $\phi$.
\begin{equation}
 t' = y\left( \frac{\tan\phi}\beta + \frac{1}{\cos\phi} - 1\right)
\end{equation}
Expanding the equation for $t'$ to second order, we get:
\begin{equation}
 t' = y\left(\frac\phi\beta + \frac{\phi^2}{2}\right)\label{eqn.9}
\end{equation}
The minimum value of $t'$ occurs at $\phi_{0}=-1/\beta$ and it is
$t'_{\rm min} = -y/2\beta^2$.  To the observer, the object first
appears at the position $\phi=-1/\beta$.  Then it appears to stretch
and split, rapidly at first, and slowing down later.

The quadratic equation~(\ref{eqn.9}) can be recast as:
\begin{equation}
  \label{eqn.q1}
  1+\frac{2\beta^2}{y}t' = \left(1+\beta\phi\right)^2
\end{equation}
which will be more useful later in the derivation.

The angular separation between the objects flying away from each other
is the difference between the roots of the quadratic
equation~(\ref{eqn.9}):
\begin{eqnarray}
 \Phi \,&=&\, \phi_1-\phi_2 \\
\,&=&\, \frac{2}{\beta}\sqrt{1+\frac{2\beta^2}{y}t'} \\
\,&=&\, \frac{2}{\beta}\left(1+\beta\phi\right)
\end{eqnarray}
making use of equation~(\ref{eqn.q1}).

Thus, we have the angular separation either in terms of the observer's
time ($\Phi(t')$) or the angular position of the object ($\Phi(\phi)$)
as illustrated in Figure~\ref{phiphi}.

\image{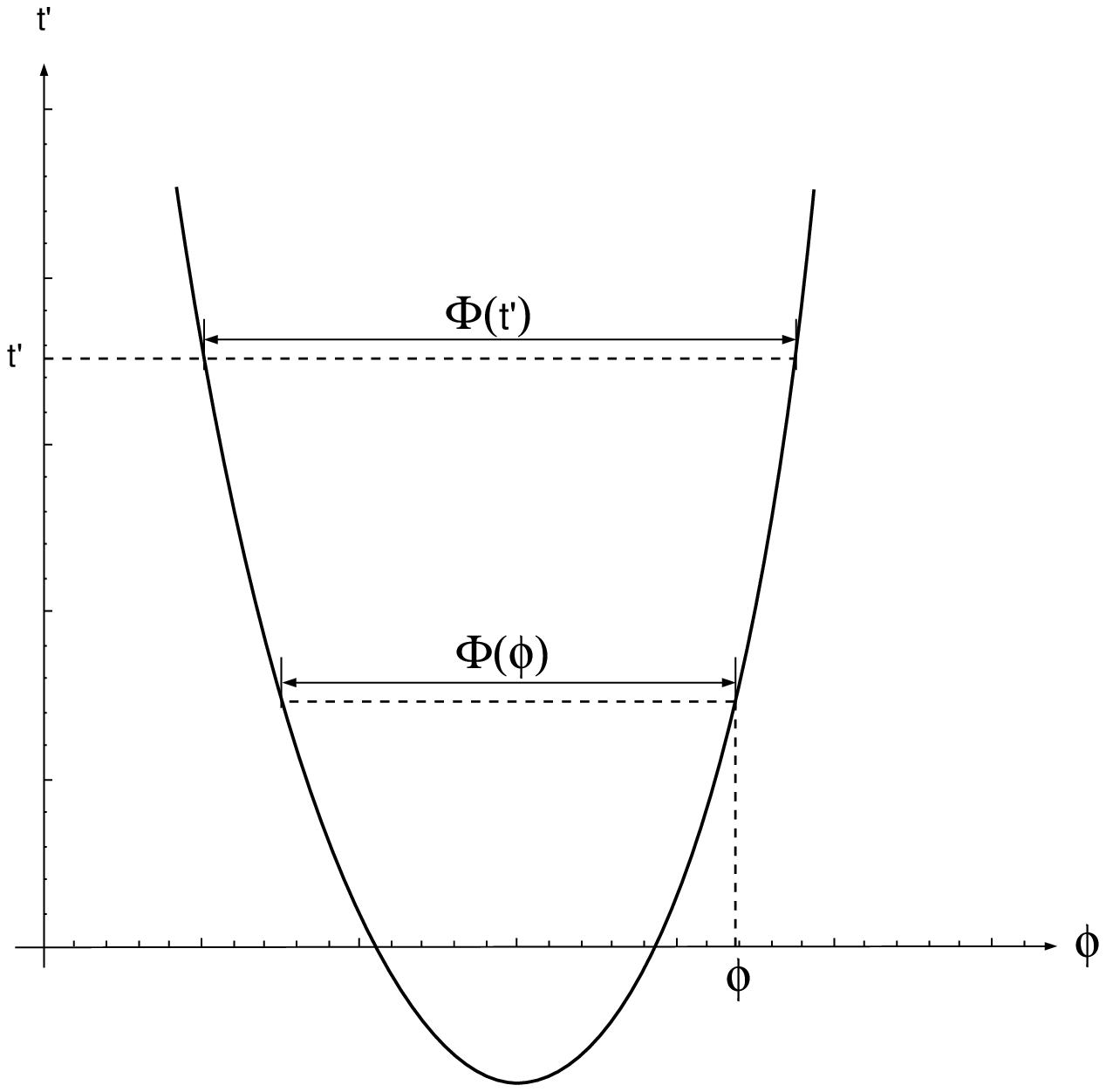}{6}{phiphi} {Illustration of how the angular separation
  is expressed either in terms of the observer's time ($\Phi(t')$) or
  the angular position of the object ($\Phi(\phi)$)}

The rate at which the angular separation occurs is:
\begin{eqnarray}
 \frac{d\Phi}{dt'} \,&=&\, \frac{2\beta}{y\sqrt{1+\frac{2\beta^2}{y}t'}}
 \\
\,&=&\, \frac{2\beta}{y\left(1+\beta\phi\right)}
\end{eqnarray}
Again, making use of equation~(\ref{eqn.q1}).

Defining the apparent age of the radio source $ t_{\rm age} = t' -
t'_{\rm min}$ and knowing $t'_{\rm min} = -y/2\beta^2$, we can write:
\begin{eqnarray}
 \frac{d\Phi}{dt'} \,&=&\, \frac{2\beta}{y\sqrt{1+\frac{2\beta^2}{y}t'}}\\
\,&=&\, \frac{2\beta}{y\sqrt{1-\frac{t'}{t_{\rm min}}}}\\
\,&=&\, \sqrt{\frac{-t_{\rm min}}{t'-t_{\rm min}}}\\
\,&=&\,\sqrt{\frac{2}{y t_{\rm age}}}
\end{eqnarray}

\subsection{Time Evolution of the Redshift}
\label{sec.dz}
As shown before in equation~(\ref{eqn.8}), the redshift $\z$ depends
on the real speed $\beta$ as:
\begin{equation}
1 \,+\, \z \,=\, \left|1 \,-\, \beta\cos\theta\right| \,=\, \left|1
  \,+\, \beta\sin\phi\right|\label{eqn.10}
\end{equation}
For any given time ($t'$) for the observer, there are two solutions
for $\phi$ and $\z$.  $\phi_1$ and $\phi_2$ lie on either side of
$\phi_0 = 1/\beta$.  For $\sin\phi > -1/\beta$, we get
\begin{equation}
 1+\z_2 = 1+\beta\sin\phi_1
\end{equation} and for $\sin\phi < -1/\beta$,
\begin{equation}
 1+\z_1 = -1 - \beta\sin\phi_2
\end{equation}
Thus, we get the difference in the redshift between the two hotspots
as:
\begin{equation}
 \Delta\z \approx 2 + \beta(\phi_1+\phi_2)
\end{equation}
We also have the mean of the solutions of the quadratic ($\phi_1$ and
$\phi_2$) equal to the position of the minimum ($\phi_0$):
\begin{equation}
\frac{\phi_1 + \phi_2}{2} = -\frac{1}{\beta}
\end{equation}
Thus $\phi_1+\phi_2 = -2/\beta$ and hence $\Delta\z = 0$.  The two
hotspots will have identical redshifts, if terms of $\phi^3$ and above
are ignored.

As shown before (see equation~(\ref{eqn.10})), the redshift $\z$
depends on the real speed $\beta$ as:
\begin{equation}
1 \,+\, \z \,=\,\left|1 \,+\, \beta\sin\phi\right| \,=\, \left|1 \,+\,
  \frac{\beta^2t}{\sqrt{\beta^2t^2 + y^2}}\right|
\end{equation}
Since we know $\z$ and $t'$ functions of $t$, we can plot their
inter-dependence parametrically.  This is shown in Fig.~\ref{zVsTp}.

It is also possible to eliminate $t$ and derive the dependence of
$1+\z$ on the apparent age of the object under consideration, $t_{\rm
  age} = t' - t_{\rm min}$.  In order to do this, we first define a
time constant $\tau = y/\beta$.  This is the time the object would
take to reach us, if it were flying directly toward us.

First, let's get an expression for $t/\tau$:
\begin{eqnarray}
 t' \,&=&\, t + \sqrt {\beta ^2 t^2  + y^2 }  - y \\
  \,&=&\, t + \beta \tau \sqrt {1 + \frac{{t^2 }}{{\tau ^2 }}}  - \beta \tau  \\
  \,&\approx&\, t + \frac{{\beta t^2 }}{{2\tau }} \\
  \Rightarrow \frac{t}{\tau } \,&=&\, \frac{{ - 1 \pm \sqrt {1 +
  \frac{{2\beta t_{\rm age} }}{\tau }} }}{\beta }
\label{eqn.ttau}
\end{eqnarray}
Note that this is valid only for $t \ll \tau$.

Now we collect the terms in $t/\tau$ in the equation for $1+\z$:
\begin{eqnarray}
 t' \,&=&\, t + \sqrt {\beta ^2 t^2  + y^2 }  - y \\
 \Rightarrow \sqrt {\beta ^2 t^2  + y^2 }  \,&=&\, t' - t + y \\
 1 + z \,&=&\, \left| {1 + \frac{{\beta ^2 t}}{{\sqrt {\beta ^2 t^2  + y^2 } }}} \right| \\
  \,&=&\, \left| {1 + \frac{{\beta ^2 t}}{{t' - t + y}}} \right| \\
  \,&=&\, \left| {1 + \frac{{\beta ^2 \frac{t}{\tau }}}{{\frac{{t_{{\rm
 age}} }}{\tau } - \frac{1}{{2\beta }} - \frac{t}{\tau } + \beta }}}
 \right|
\label{eqn.zz}
 \end{eqnarray}
As expected, the time variables always appear as ratios like $t/\tau$,
giving confidence that our choice of the characteristic time scale is
probably right.

Finally, we can substitute $t/\tau$ from equation~(\ref{eqn.ttau}) in
equation~(\ref{eqn.zz}) to obtain:
\begin{equation}
1 + \z = \left| {1 + \frac{{\beta ^2 \left( { - \tau  \pm \sqrt
            {2 \beta  t_{{\rm age}}} } \right)}}{{\beta t_{{\rm age}}
        + \tau /2 \mp \sqrt {2 \beta t_{{\rm age}}}  + \beta ^2 \tau
      }}} \right|
\end{equation}

\subsection{Time Evolution of the Object Size}
\label{sec.dx}
Fig.~\ref{PhiVsTp} shows the apparent positions ($\phi$) and the size
of the superluminal object as the observer sees it, as a function of
the observer's time ($t'$).  Fig.~\ref{zVsTp} is a similar time
evolution of the redshift ($\z$).  In this section, we describe how
these two plots are created.  It is easiest to express the quantities
parametrically as a function of the real time $t$. Referring to
Fig.~\ref{dd}, we write,
\begin{eqnarray}
  x &\,=\,& \beta t \\
t' &\,=\,& t + \sqrt{\beta^2t^2 + y^2} - y\\
\sin\phi &\,=\,& \frac{x}{\sqrt{x^2 + y^2}} = \frac{\beta t}{\sqrt{\beta^2t^2 + y^2}}
\end{eqnarray}
The solid parabola in Fig.~\ref{PhiVsTp} is $\phi$ vs. $t'$ from these
equations as $t$ is varied between $-40$ and $20$ years, with $y =
1\,000\,000$ light years and $\beta = 300$.

In order to get the variation of the size of the object (the shaded
region in Fig.~\ref{PhiVsTp}), we assume a diameter $d = 500$ light
years.
\begin{eqnarray}
t'_\pm \,&=&\, t + \sqrt{\left(\beta t \pm \frac{d}{2}\right)^2 + y^2} - y\\
\sin\phi_\pm \,&=&\, \frac{\beta t \pm \frac{d}{2}}{\sqrt{\left(\beta t \pm
    \frac{d}{2}\right)^2 + y^2}}
\end{eqnarray}
The boundaries of the shaded region are given by $\phi_+$ vs. $t'_+$
and $\phi_-$ vs. $t'_-$.

\section{Perceived Properties of the Universe}
\label{sec.universe}
It can be shown that the apparent expansion of the universe at
strictly subluminal speed is also an artifact of our perception of
superluminal motion.  The apparent recessional speed is the
longitudinal component of $\beta'$ is $\beta'_\parallel = \beta'
\cos\theta$.  From equation~(\ref{eqn.1}), we can see that
\begin{equation}
\lim_{\beta\to\pm\infty} \beta'_\parallel \,=\, -1
\end{equation}
The apparent recessional speed (which can be measured using redshifts)
tends to $c$ (or, $\,\beta'_\parallel\,\to\,-1$), when the real speed
is highly superluminal.  This limit is independent of the actual
direction of motion of the object $\theta$.  Thus, whether a
superluminal object is receding or approaching (or, in fact, moving in
any other direction), the appearance from our perspective would be an
object receding roughly at the speed of light.  This appearance of all
(possibly superluminal) objects receding from us at strictly
subluminal speeds is an artifact of our perception, rather than the
true nature of the universe.

Note that the equation for $1+\z$ has a limiting value of the real
speed $|1+\beta|$ as large angles $\phi$.  Thus, if we picture our
universe as a large number of superluminal or hyperluminal objects
moving around in random directions, there will be a significant amount
of low frequency isotropic electromagnetic radiation.  The spectrum of
this cosmic microwave background can be directly translated to a
velocity distribution of the celestial objects.  The spatial asymmetry
in the background radiation corresponds to a temporal evolution of the
superluminal objects.  Thus, even the cosmic microwave background
radiation (considered one of the strongest arguments for the big bang
model) can be accommodated in our model.

While working out various kinematic properties of superluminal
objects, we noted that there is an {\em apparent} violation of
causality when the superluminal object is approaching us.  We also saw
that a receding object can never {\em appear} to be going faster than
the speed of light, even if the real speed is superluminal. For a
receding object (even subluminal ones), there is a contraction of
object size along the direction of motion and a time dilation effect.
All these effects are due to the light travel time effect, but are
remarkably similar to the theory of special relativity.  However, the
light travel time effect is currently assumed to apply on a
space--time that obeys SR.  It may be that there is a deeper structure
to the space--time, of which SR is only our perception, filtered
through the light travel time effect.  By treating light travel time
effect as an optical illusion to be applied on an SR--like
space--time, we may be double counting the effects.

\bibliographystyle{elsart-harv}
\bibliography{refs}

\end{document}